\documentclass[a4paper]{ESASPCS13Style}
\usepackage{epsfig, amsmath, pifont}

    \def\etal{${\rm \hspace*{0.8ex}et\hspace*{0.7ex}al.\hspace*{0.5ex}}$}
    \def\plus{${\rm \hspace*{0.7ex}\&\hspace*{0.7ex}}$}
    \def\ie{i.\,e.\ }
    \def\tmix{\tau_{\rm mix}}
    \def\aquer{\langle a\rangle}
\begin{document}
\sloppy

\title{Rain and clouds in brown dwarf atmospheres:\newline
A coupled problem from small to large}

\author{Ch. Helling\inst{1,2} \and P. Woitke\inst{1}} 
\institute{Sterrewacht Leiden, P.O. Box 9513, 2300 RA Leiden, The Netherlands
  \and  Zentrum f\"ur Astronomie \& Astrophysik, TU Berlin, Hardenbergstra{\ss}e 36, 10623 Berlin, Germany}

\maketitle 

\begin{abstract}

The large scale structure of a brown dwarf atmosphere is determined by
an interplay of convection, radiation, dust formation, and
gravitational settling, which possibly provides an explanation for
the observed variability. The result is an element depletion of the
dust forming regions and an element enrichment of the dust evaporating
sites. The formation of dust cloud structures in substellar
atmospheres is demonstrated based on a consistent theoretical
description of dust formation and destruction, gravitational settling,
and element depletion including the effect of convective overshoot.

Since the viscosity is small in brown dwarf atmospheres, the convection
creates a turbulent environment with fluctuations of all thermodynamic
state variables on a wide range of spatial scales.  Hence, the
classical turbulent closure problem needs to be tackled in connection
with dust formation  in substellar objects, because a complete
three-dimensional and time-dependent solution of the model equations
is simply not possible.  Structure formation may be
seeded on the smallest scales, when chemical processes are
involved.  In order to understand the interaction of turbulence and
dust formation, we have performed investigations of the smallest scale
regimes in 1D and in 2D in order to identify the governing processes
of the unresolved scale regime.

\smallskip
{Stars: substellar -- Planets: atmosphere -- dust, turbulence }
\end{abstract}

\section{Introduction}
The large scale structure of large parts of a brown dwarf atmosphere
is influenced by the life cycle of dust (\cite{h2003,wh2003}) which
eventually governs their spectral appearance. This life cycle of dust
is a consequence of an interplay of convection, dust formation
(nucleation and growth), evaporation and gravitational settling
(\cite{holks2001b,wh2003}) thereby providing a likely cause for the
observed non-periodic atmospheric variability (e.g. \cite{bjm1999};\\
\cite{cot2002,k2004}). Dust is the most active chemical component in
the stellar atmosphere fluid because it is {\it i)} the main opacity
carrier in substellar atmospheres once it has formed, {\it ii)} the
most active chemical component which consumes condensable elements
from the gas phase, and hence, {\it iii)} it changes
the atmospheric $(\rho, T)$ structure.

Every convectively unstable gas will be turbulent providing a source
of additional small scale variability. Because a complete
three-dimensional and time-dependent solution of the radiative
transfer and the fluid dynamics including a complex chemistry is
simply not possible yet, the turbulent closure problem must be tackled
in connection with dust formation and element conservation in
substellar objects. A numeric approach to such challenges is the Large
Eddy Simulation (LES) method in which the range of resolvable scales
is numerically treated and the unresolved, or residual scales are
represented by a subgrid model.

Anticipating for instance the IR Spritzer observations with a spectral
range up to $\approx 30\mu$m (see \cite{r2004}) which luckily is
typical for dust, we followed two paths in our investigations:
\begin{enumerate}
\item The theoretical background of gravitational settling (rain out)
is studied on the base of extended dust moment equations which allow
for a simultaneous modelling of dust nucleation, growth/evaporation and
size dependent drift (Sect.~\ref{sec:lsr}).
\item We have investigated the small scale regime in 1D and in 2D to
identify the governing processes, and to understand the interaction
and the feedback mechanisms of turbulence and dust formation
(Sect.~\ref{sec:ssr}).
\end{enumerate}

\begin{figure}
\centerline{\epsfig{file=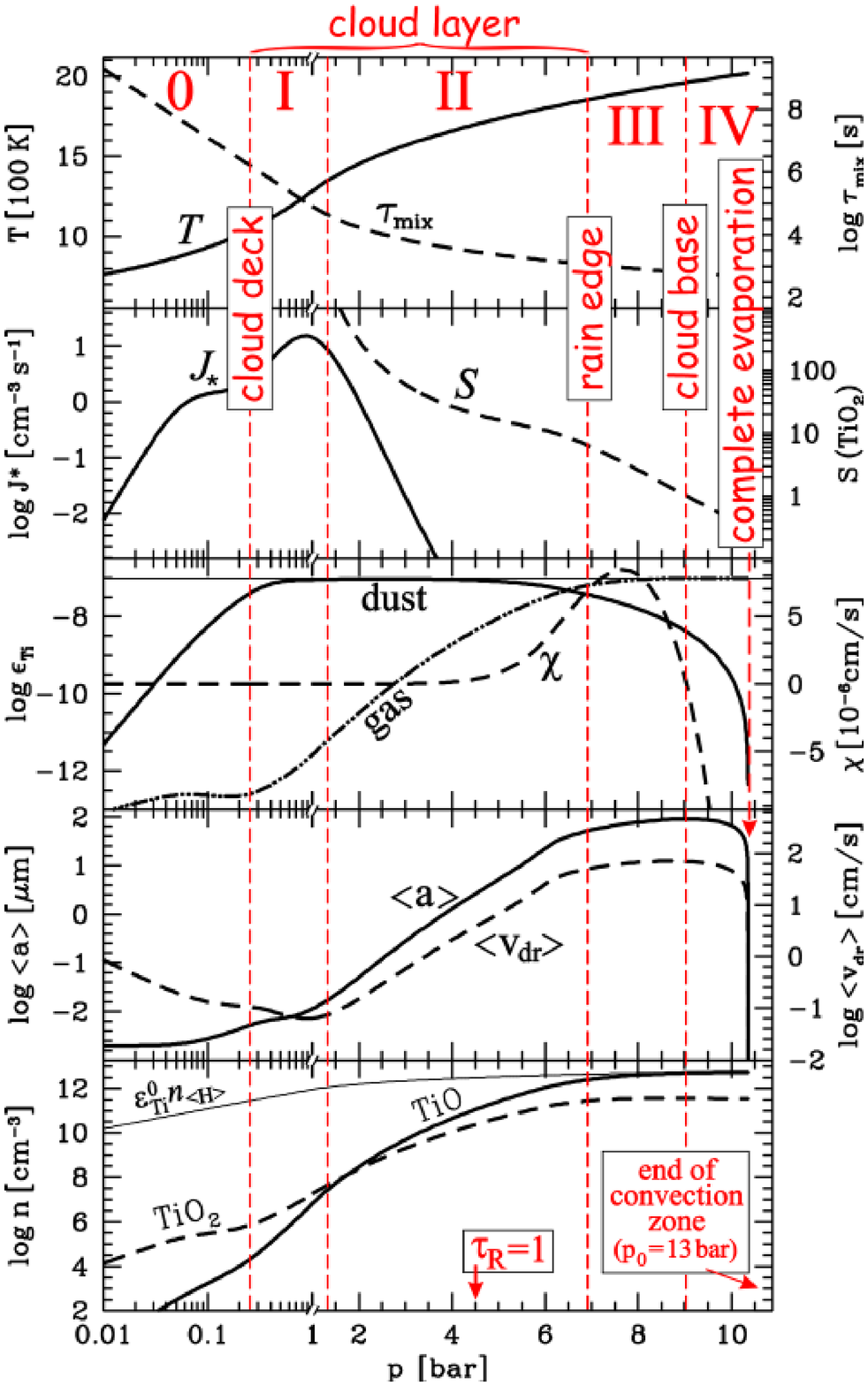, scale=0.5}}
\centerline{\epsfig{file=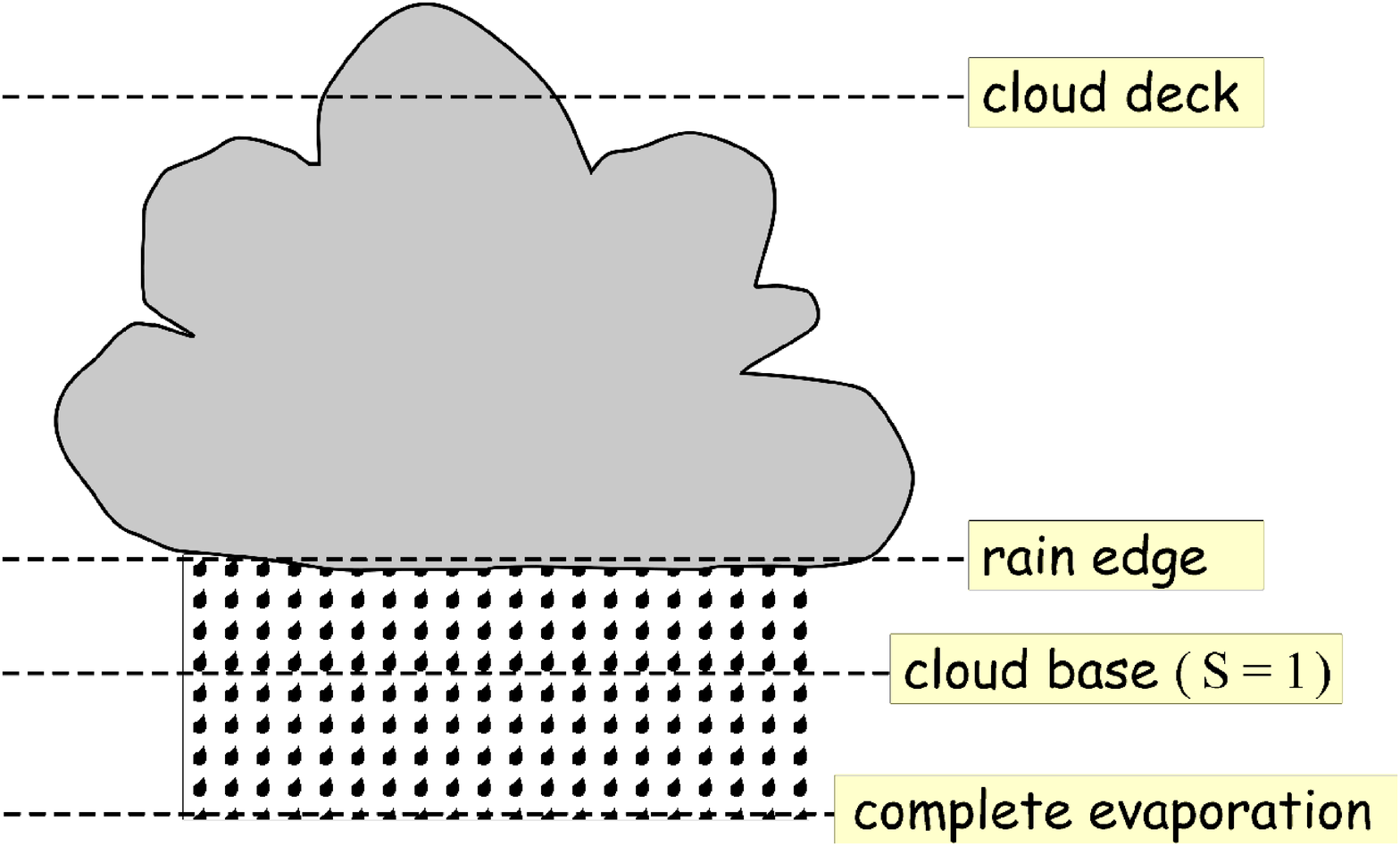, scale=0.15}}
\caption{The substellar, large-scale, quasi-static dust cloud
structure is merely stratified by the dust formation processes inside
the cloud. Eventually, the dust particles rain out, evaporate and
thereby element enrich deeper atmospheric regions. Fresh, uncondensed
material is mixed up by convection and subsequent convective
overshoot. ($T_{\rm eff}=$1400K, $g = 10^5 $cm\,s$^{-1}$)}
\label{fig1}
\end{figure}

\section{
The large-scale structure
}
\label{sec:lsr}
\cite{wh2003} have extended the theory of dust formation developed by
\cite{gs86} into the substellar regime. The dust formation can be
treated as a two step process because of time-scale arguments: first
seed particles form (nucleation) which subsequently grow to
macroscopic sizes. \cite{gs86,gs88} developed a system of dust moment
equations which have the character of conservation equations with
source terms due to nucleation and growth/evaporation. By taking into
account additionally the size dependent drift -- particularly
important in substellar atmospheres --, simultaneous with nucleation
and growth/evaporation, a time-dependent system of extended dust
moment equation results which has to be solved consistently with the
element conservation equations. A handsome set of static
($\boldsymbol{v}_{\rm gas}=0$), stationary ($\partial L_{\rm
j}/\partial t =0$) dust moment equations including auxiliary
conditions for element conservation has been derived in \cite{wh2004}.
Its solution for TiO$_2$-grains on top of a prescribed substellar
model structure $(p, T)$ has been demonstrated (for model details
please consult \cite{wh2004}).  Figure~\ref{fig1} (top) shows the
calculated stratification which results from the interplay between
convective up-mixing, dust formation, and drift. The up-mixing is
modelled as a convective overshoot with exponentially decreasing
efficiency with hight above the convective active zone.

The large-scale quasi-static cloud distribution in a brown dwarf
atmosphere results from two counteracting processes: The upward mixing
of condensable elements by convective overshoot and the downward rain
out. The distribution of dust inside and close to the cloud is merely
determined by the sensitive dependence of the dust formation processes
(nucleation, growth/evaporation) on the local temperature ($T$), the
local gas density ($\rho$), the local element abundances
($\epsilon_{\rm i}$), and the local mixing time scale.  The bottom sketch
in Fig.~\ref{fig1}  visualises the close resemblance of
terrestrial expectations of clouds and rain.

\subsection{Atmospheric dust component}\label{ss:dust}
The study of the dimensionless dust moment equations  for
typical substellar atmosphere conditions reveals a hierarchical
appearance of the source terms nucleation $\longrightarrow$
growth/evaporation $\longrightarrow$ drift (\cite{wh2003}). Nucleation
(seed particle formation) is the dominant source term (region I in
Fig.~\ref{fig1}) if the local thermodynamic conditions allow for
it. Only if nucleation becomes inefficient or impossible, the
growth/evaporation term dominates (regions II and IV in
Fig.~\ref{fig1}). Only if nucleation and growth are both inefficient,
the drift term will govern the equations solution (region III and
partly also regions II, IV in Fig.~\ref{fig1}).

\paragraph*{\underline{0. Dust-poor depleted gas:}\,\,} High
above the convection zone, the mixing time-scale is large and the
elemental replenishment of the gas is too slow to allow for
considerable amounts of dust to be present in the atmosphere. The few
particles forming here are very small which is, however, sufficient to
make $\tau_{\rm sink} < \tmix$, which causes these atmospheric layers
to become dust-poor.  The gas phase is strongly depleted in
condensible elements.

\paragraph*{\underline{I. Region of efficient nucleation:}\,\,}
Nucleation takes place\newline mainly in the upper parts of the cloud
layer, where the temperatures are sufficiently low and the elemental
replenishment by mixing is sufficiently effective. Although the gas is
strongly depleted in heavy elements in these layers, it is
nevertheless highly supersaturated ($S\!>\!1000$) such that
homogeneous nucleation can take place efficiently.  The dust grains
remain small $\langle a\rangle\!<\!0.01\,\mu$m. The mean drift
velocities are even smaller than in region~0 because of the higher gas
densities.

\paragraph*{\underline{II. Dust growth region:}\,\,} 
With the inward  increasing temperature, the supersaturation ratio $S$
decreases exponentially which leads to a drastic decrease of the
nucleation rate $J_\star$. Consequently, nucleation becomes
unimportant at some point, \ie {\it the in-situ formation of dust
  particles becomes inefficient in region~II}.  Here, the condensible
elements mixed up by convective overshoot are mainly consumed by the
growth of already existing particles, which have formed in region~I
and have drifted into region~II. The gas is still strongly
supersaturated $S\!\gg\!1$, indicating that the growth process remains
incomplete, \ie the condensible elements provided by the mixing are
not exhaustively consumed by growth.

\paragraph*{\underline{III. Drift dominated region:}\,\,}
With the increasing mean size of the dust particles, the total surface
decreases which makes the consumption of condensible elements from the
gas phase less effective.  At the same time the drift velocities
increase. When the dust particles have reached a certain critical
size, $\aquer_{\rm cr}\!\approx\!15\,\mu$m to $50\,\mu$m, the drift
becomes more important than the growth, and the qualitative behaviour
of the dust component changes ({\it rain edge}).  Although the gas is
still supersaturated $S\!\approx\!1\,...\,10$, the in-situ formation
of dust grains is as ineffective as in region~II.  The depletion of
the gas phase vanishes in region~III, \ie the gas abundances approach
about the solar values.  The grains reach their maximum radii at the
lower boundary of region~III ({\it cloud base}):
$\aquer\approx\!30\,\mu$m to $400\,\mu$m.

\paragraph*{\underline{IV. Evaporating grains:}\,\,}
The gravitationally settling dust particles finally cross the cloud
base and sink into the under-saturated gas situated below, where
$S\!<\!1$ and $\chi_{\rm lKn}^{\rm net}\!<\!0$. Here, the dust grains
raining in from above evaporate thermally. The evaporation of these
dust particles, however, does not take place instantaneously, but
produces a spatially extended evaporation region~IV with a thickness
of about 1\,km for the model atmosphere shown here.

\subsection{Atmospheric gas component}\label{ss:gas}

Beside the general cloud structure (Figs.~\ref{fig1}) and details
about the atmospheric dust component like rate of nucleation $J_*$
(solid line, 2nd panel Fig.~\ref{fig1}, top), mean particle sizes and
mean drift velocities (both 4th panel Fig.~\ref{fig1}, top), we find
also detailed information about the molecule abundance in the gas
phase (5th panel Fig.~\ref{fig1}, top). The remaining gas composition
depends on the dust formation processes and  follows the
atmosphere structuring by the dust as suggested in Sect.~\ref{ss:dust}.

\paragraph*{\underline{0, I, II. Regions of metal deficiency:}\,\,}
The gas phase is\newline strongly depleted by those elements involved
into the dust formation process (merely Ti in the case depicted in
Fig.~\ref{fig1}). This metal deficiency amounts for instance to about
6 orders of magnitude in the case of TiO$_2$ in the visible outer
region 0.

This metal deficiency decreases with decreasing height and its
observational relevance will depend on where the $\tau=1$--level is
reached. The model atmosphere adopted here would appear strongly metal
deficient. 

\paragraph*{\underline{III. Region of nearly solar metalicity:}\,\,}
The gas phase has approximately solar abundances since it is strongly
mixed with gas from the deep interior and the depletion by the growth
of the few large particles drifting through is has only little
influence.

\paragraph*{\underline{IV. Region of metal enrichment:}\,\,}
The size dependent evaporation of the inward drifting dust grains
causes a height dependent element enrichment of the inner, hot and
dense atmospheric regions. It might, however, lay below the
$\tau=1$--level and may therefore be only visible for rather warm
substellar objects. 

\begin{figure}
\hspace*{-0.5cm}\epsfig{file=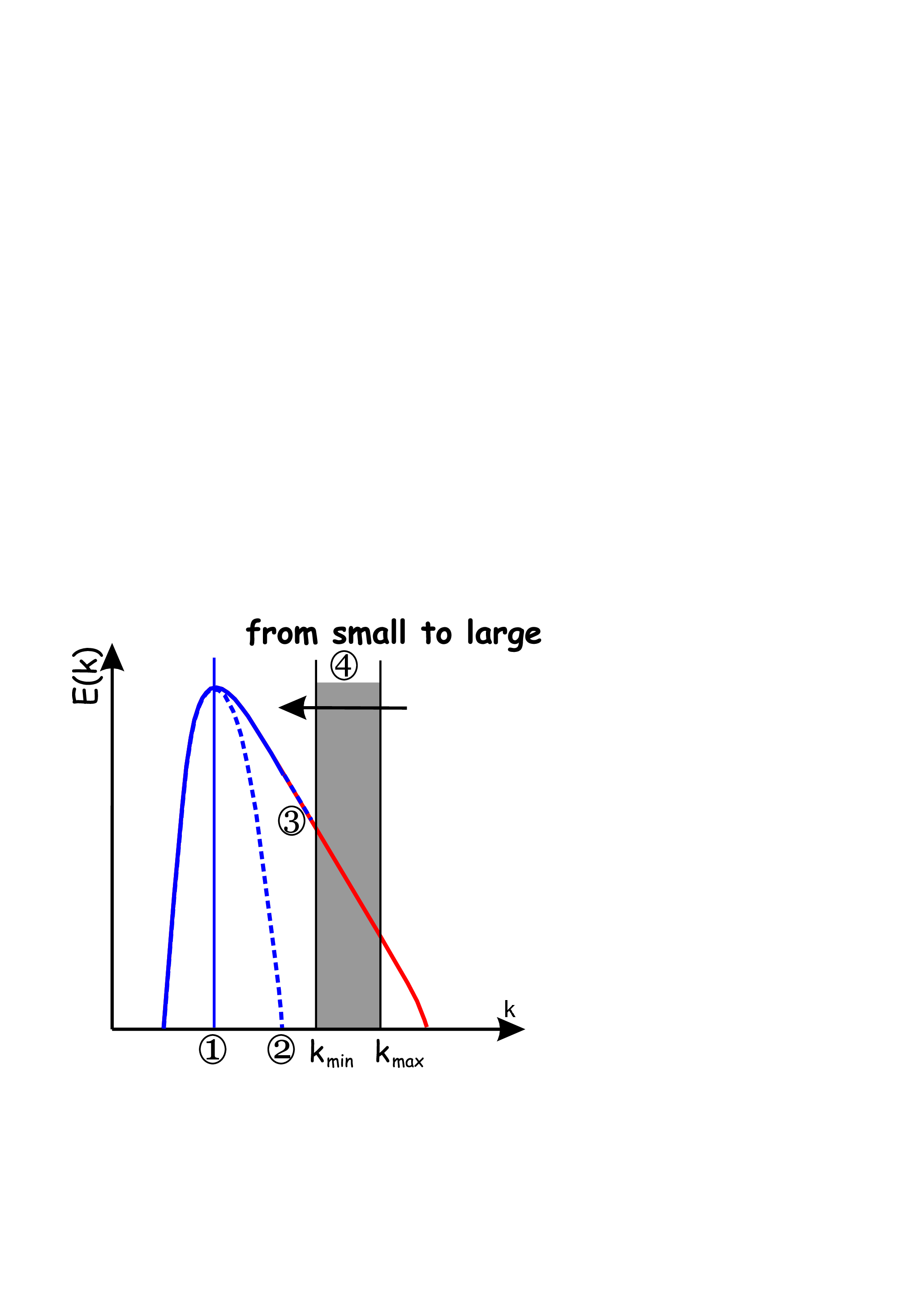, scale=0.7}
\vspace*{-3.5cm}
\caption{Model approaches to the Kolmogoroff scale spectrum for
turbulent media: \ding{172} mixing length theory (MLT), \ding{173}
artificial viscosity, \ding{174} large eddy simulation (LES)
\ding{175} this work. }
\label{fig2}
\end{figure}

\section{The small-scale regime: Cloud formation}
\label{sec:ssr}


If the inertia of the atmospheric gas is larger than
its friction, \ie large Reynolds number, turbulence will develop if a
driving mechanism is present. Substellar atmospheres seem to provide
both: high Reynolds numbers and convection as driving
mechanism. Consequently, simulations of brown dwarf and planetary
atmospheres have to deal with the multi-scale problem as it is
characteristic for turbulent systems.  In classic static model
atmospheres the mixing length theory is used to describe the
multi-scale problem of turbulent convection. Only one -- the most
energy containing scale of the system $k_{\rm MLT}$ -- is used to
model the whole scale spectrum (Fig.~\ref{fig2}\, \ding{172}).

In simulations of dynamic model atmospheres more advanced methods have
been used, like the artificial viscosity (Fig.~\ref{fig2}\,
\ding{173}) which damps fluctuations such that no energy is contained
in the smallest but still resolvable scale ($k_{\rm resol}$). The
large eddy simulation tries to use information from the edge of the
resolvable regime to model the residual, unresolved scales
(Fig.~\ref{fig2}\, \ding{174}). Obviously, knowledge needs to be
accumulated about this small-scale regime. A hierarchical
investigation from small to large scale has been started by
\cite{holks2001b} as sketched by \ding{175} in Fig.~\ref{fig2}.  The
basic idea is that convective elements do collide and thereby create a
whole spectrum of disturbances, \ie turbulent fluctuations.
\begin{figure}
\begin{center}
\epsfig{file=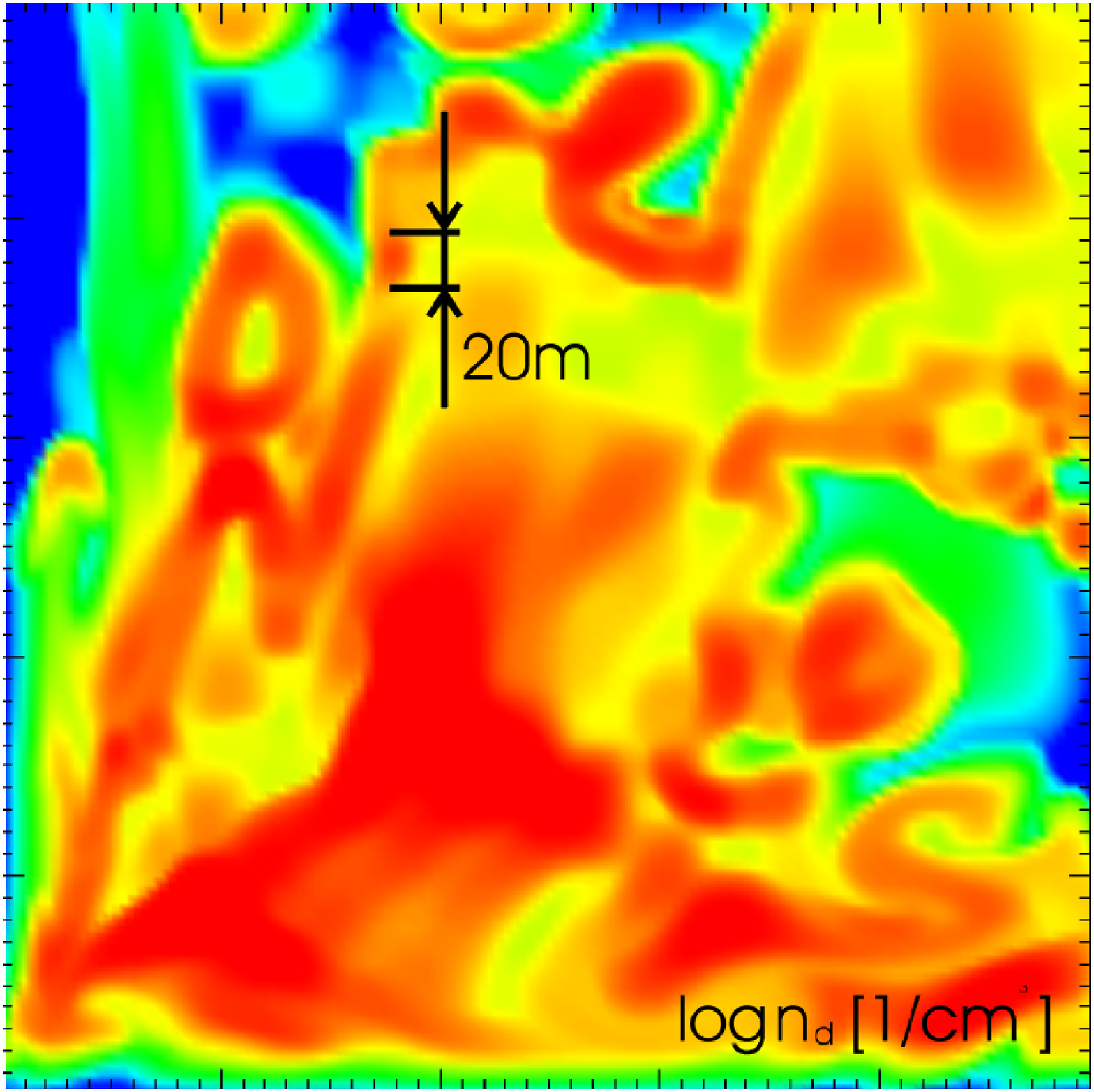, scale=0.25}\\[0.3cm]
\epsfig{file=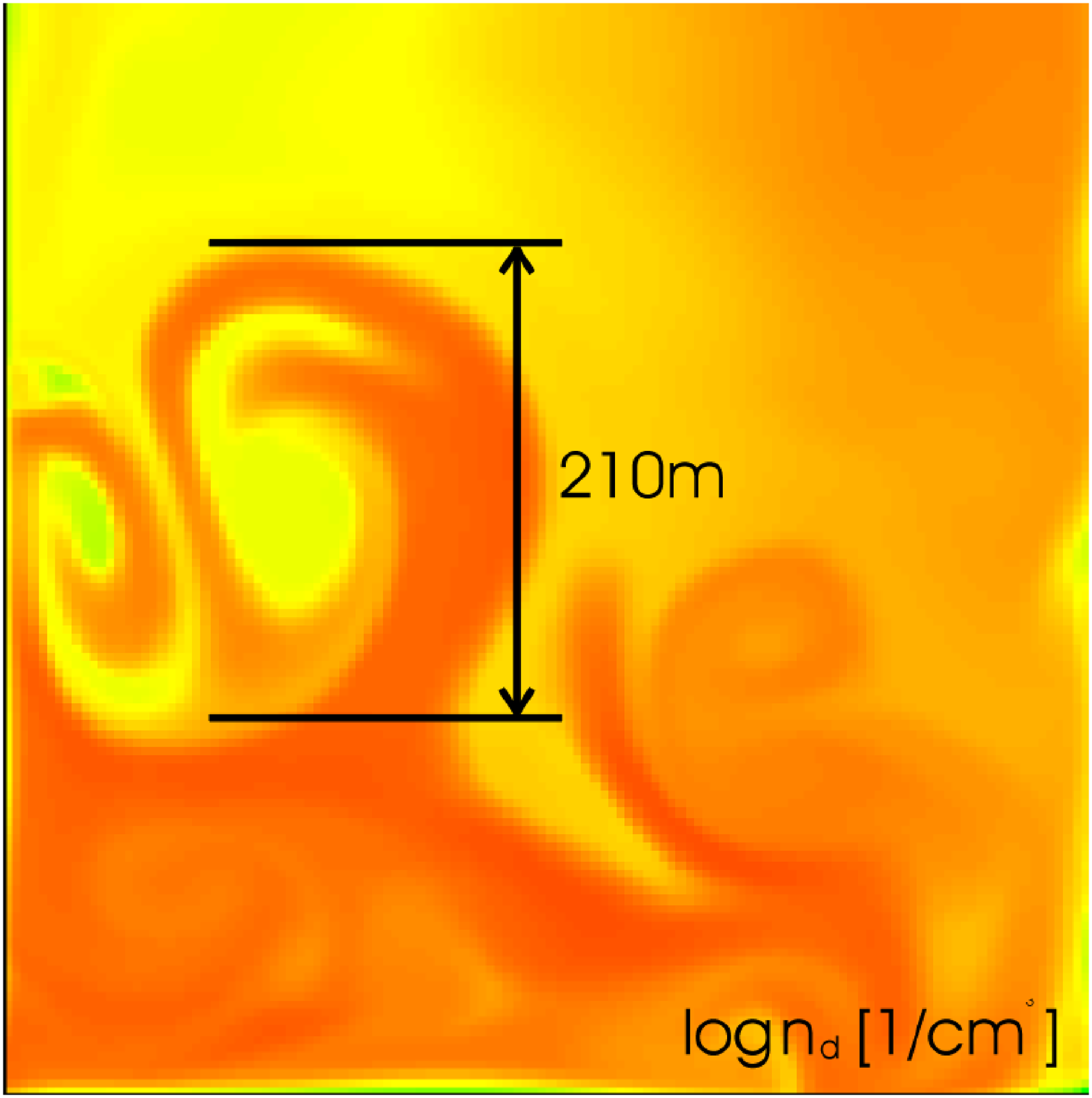, scale=0.251}
\end{center}
\caption{Turbulent dust formation in the macroscopic regime (top:
$t=0.436$\,s, bottom: $t=8.3$\,s). The results are depicted for the
number of dust particles (red/dark grey: $\log n_{\rm d}=10$ [with
n$_{\rm d}$ in cm$^{-3}$], yellow/light: $\log n_{\rm d}=8$ cm$^{-3}$,
blue/black: $\log n_{\rm d}= 0$).
\newline Model parameter: $T_{\rm ref}=2100\,$K, $T_{\rm RE}=1980\,$K,
$\rho_{\rm ref}= 3.16\,10^{-4}$g\,cm$^{-3}$, $t_{\rm ref}=0.3\,$s,
$l_{\rm ref}=10^5$cm, $\Delta x= 3.94\,10^2$cm, $N_x\times N_y = 128
\times 128$, N$_{\rm k}=500$.}
\label{fig3}
\end{figure}
Fluctuations carry expansion waves which can initiate dust formation
by a local drop of the temperature below the nucleation threshold. If
thermal stability is assured, these seed particles grow to macroscopic
sizes and start to radiate thermally. The result is a reinforced
cooling of the gas/dust mixture in the small scale regime. A feedback
loop sets in which assures more dust to form until all condensible
material is consumed from the gas phase. The dust complex has then
reached a steady state (compare Fig.~3 in \cite{holks2001b}).

Relying on these results, \cite{hkwns2004} have adopted the concept
of driven turbulence where a whole spectrum of superimposed waves
constantly enters and thereby disturbs an initially dust free gas
inside a test volume. Figure~\ref{fig3} demonstrates the principle
findings: The dust formation process is initiated on the smallest
length scales resulting in a very inhomogeneous distribution of
dust. Similar results were obtained based on asymptotic analysis of
the nucleation rate in Fig.~2 in \cite{holks2001b}.  The smallest
observable dust structure in the present Fig.~\ref{fig3} (top) is
$\approx 20$m after $t=0.436$\,s with the smallest resolved scale
being $\Delta x \approx 4\,$m. These small scale heterogeneity lasts
not very long. Ongoing hydrodynamic disturbances cause the dust
structures to gather in larger and larger cloud-like structures
causing a (mesoscopic) large-scale heterogeneity. Lanes and curls
appear and disappear during these dynamic process of
gathering. Structures of the size of $\approx 200$m have developed in
the presented simulation after $\approx 8$s (Fig.~\ref{fig3},
bottom). Note that the appearance of curls indicates the development
of vorticity which is typical for turbulent systems. During the time
of active dust formation, sites with high vorticity appear and the
question may be raised if the formation of dust can cause
turbulence. The answer seems to be positive since the appearance of
dust in a system has a strong feedback onto the system's temperature
and the velocity field, \ie the thermal and the dynamic structure of
the atmosphere.
\begin{figure}
\epsfig{file=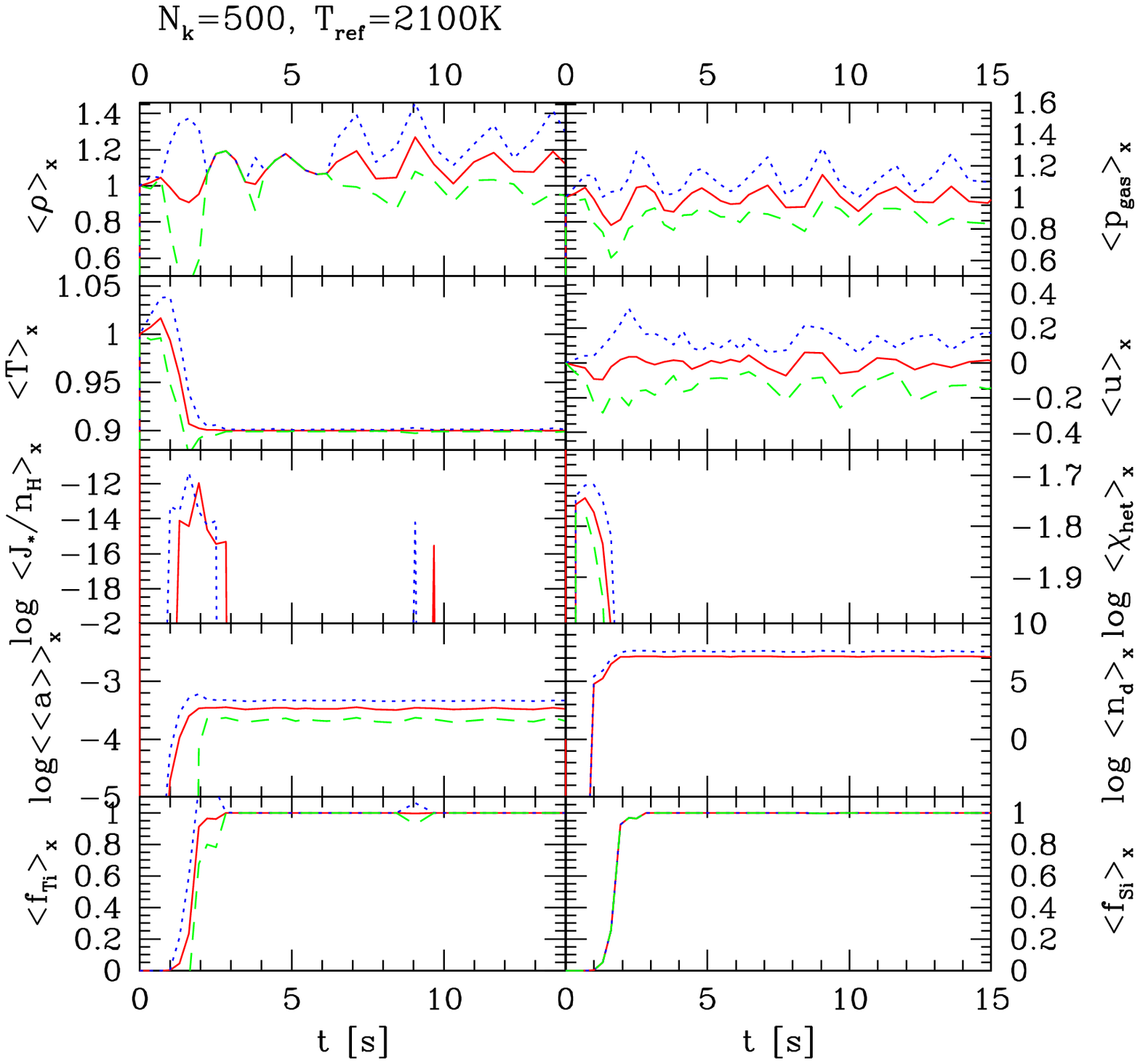, scale=0.5}
\caption{The space-means $\langle\alpha(t)\rangle_x$ (solid/red) plus/minus
the apparent standard deviations $\sigma_{\rm N_x-1}^{\alpha}(t)$
(dotted/dashed) as a function of time for the same parameter like
Fig~\ref{fig3} but for t$_{\rm ref}=3.08$\,s and N$_{\rm
x}=500$. ($\langle\alpha(t)\rangle_x+\sigma_{\rm N_x-1}^{\alpha}(t)$ -
dotted/blue; $\langle\alpha(t)\rangle_x-\sigma_{\rm
N_x-1}^{\alpha}(t)$ - dashed/green).}
\label{fig4}
\end{figure}
This idea is supported by the results of 1D simulations of which one
example is shown in Fig.~\ref{fig4} where means with standard deviations
are depicted.  The apparent standard deviation is largest in the
period of most active dust formation, \ie between 0.05\,s and 3\,s,
independent of the initial reference temperature. Almost no variations
are present in the dust quantities if the dust complex has reached its
steady state which is much in contrast with the hydrodynamic
quantities (top four panels in Fig.~\ref{fig4}) which are continuously
perturbed by the turbulence driving mechanism.

\paragraph{Optical depth:}

Adopting Rosseland mean gas opacity typical for the hot, inner layers
of a brown dwarf atmosphere and the Rosseland mean dust opacity for
astronomical silicates, the variability of the total opacity and the
space mean optical depth for the size of our test volume (500\,m) has
been investigated.  For the simulation depicted in Fig.~\ref{fig4} the
dust and the gas opacities differ by about 1.5 orders of
magnitude. The dust opacities vary by about 0.5 magnitudes, whereas
the gas opacity by only about 0.2 order of magnitudes. This is
independent of the characteristic time scale of the system, \ie the
large-scale Mach number of the initial configuration.  The optical
depth increases by a factor of 10 when the dust forms, in accordance
with the opacity increase. There is, however, a delay between the
onset of dust formation and the time when the maximum optical depth is
reached because the dust formation sets in earlier than a cloud
becomes optically thick. This delay is sensitive to the characteristic
time scale of the system.

The gas/dust mixture will only be optically thick for the case of
non-transparent dust particles depending also on the wavelength
considered. Glassy grains efficiently absorb in the far IR
($\lambda\ga 10\mu$m) and one might, depending on the observed
wavelength, detect typical dust features and possibly even the rapid
formation events which, however, may need to evolve on macroscopic
scales to be observable.

Turbulent fluctuations do not seem to produce considerable variations
in view of the space mean of the optical depth in the long term (here
$t>4$\,s) if astronomical silicates are assumed as typical dust
opacity carriers, though turbulent fluctuations are essential for the
initiation of dust formation and eventually for the origin of large
scale structures.

\paragraph{Dependence on Mach number:}
The use of dimensionless equations allows to describe the character of
the model equations and their solution by a set of characteristic
numbers. For instance, the Mach number = 1 suggest a sound-like flow
while Mach number $\ll 1$ implies a quasi-static fluid field.  This
concerns only the mean behaviour of the reactive fluid. Deviations
from this can occur and a local rather than a global Mach number may
occasionally be considered. The Mach number of the local turbulent
fluctuations is always sub-sonic due to the underlying model of driven
turbulence (see \cite{hkwns2004}) in our simulations, \ie M$_{\rm
turb}<1$.  The global Mach number describes the turbulent energy input
on the largest resolvable scale and has an indirect influence on
M$_{\rm turb}$.  However, the dust formation is a local process, and
its onset is delayed in case of low global Mach number. The overall
pattern formation remains very much alike but needs longer times to
evolve for smaller Mach numbers.

\section{Challenges}\label{sec:cha}

Presently, two schools exist for modelling the large-scale structures
of substellar atmospheres. Both aim on the understanding of the
physiochemical processes being responsible for the detected spectral
appearance and the photometric variability.

 a) The classic static model atmosphere includes -- beside the
assumption of hydrostatic equilibrium -- an elaborate frequency-dependent
(static) radiative transfer treatment assuring flux conservation also
with regard to the convective flux treated in the mixing length
approximation. The solution of the chemical equilibrium for hundreds
of gaseous - and if needed also for solid - species closes the system
of model equations.

 b) Multi-dimensional modelling originating from solar and engineering
research. The complete time-dependent hydrodynamic equations are
solved including energy conservation. The time-dependence and
multi-dimensionality is payed by either a simplified radiative
transfer treatment (\cite{lah2002}).  In addition, the treatment of
the gas phase chemistry is very limited.  Only \cite{holks2001b} did
include a time-dependent model for dust formation (see
Sect.~\ref{sec:lsr}).

It seems tempting to combine the advantage of the one
(e.g. frequency-dependent radiative transfer for spectrum
interpretation) with the advantage of the other (e.g. hydrodynamic
cloud formation). 

What do we need to achieve such a goal?

\paragraph{\underline{Hydrodynamics: Convection \& Turbulence}}
A complete hydrodynamic modelling of a substellar atmosphere will need
to solve the multi-scale problem inherent to any turbulent
media. Hence, we need a model able to describe the large scale
hydrodynamic structures commonly ascribed to convection. The model
must simultaneously describe the small-scale turbulent fluid
field. One way to do so is to apply the method of the Large Eddy
Simulation (LES) where the spatial and the associated time scale are
resolved down to a cut-off length. The scales smaller than that must
be represented by some appropriate subgrid model.

A further challenge is the simulation of the physical system close to
a quasi-static situation.  Numeric truncation errors may cause unwanted
spurious effects which can accumulate in the course of the simulation.

\paragraph{\underline{Hydrodynamics: Rotation}} Brown dwarfs are 
shown to be fast rotators (see Scholz \& Eisl\"offel this conference)
and we know from Jupiter that rotation shapes its initially
intermittent atmosphere.  Rotation will introduce a day/night
behaviour and tides on the scale of the entire globe. It is, hence, a
large-scale process and a globe-sized simulation will be needed. So
far, only rotation free simulations are available for brown dwarfs.

\paragraph{\underline{Radiative transfer}}
For spectrum interpretation and for a correct atmospheric temperature -
density structure, the photon - matter interaction has to be
modelled. Huge data bases with millions of lines for the various
molecules of interest (H$_2$O, TiO, TiC, TiH, CO, SiO, VO, CH$_4$, CaH,
CrH, FeH, H$_3$O, NH$_3$, PH$_3$, CIA H$_2$) and others, which
influence the thermal structure but not show up in spectral features,
are available and under construction. Obeying observational evidence,
atomic data are needed as well as good models for line profiles (K~I,
Na~I, Ca~II; Rb~I, Cs~I).

Multi-dimensional and time-dependent simulations are not able to
handle such amounts of data which makes the 1D models the best and
only tool for detailed spectrum interpretation also in the far
future. However, the radiative transfer in spatially inhomogeneous
media may need further attention also in substellar objects since it
can support structure formation in dusty media where e.g. shadow
casting is a major effects (see question by Steinacker).

\paragraph{\underline{Gas phase chemistry}}
So far, time-dependent models were presented which include only a very
simple model for the chemical gas phase composition of the dust
forming gas. Our work in progress allows to calculate the chemical
equilibrium among 33 molecules and the conservation in space and time
of 10 elements which is sufficient for the study of heterogeneous dust
formation.

It may be necessary to study chemical non-equi\-librium effects
influencing the composition of the gas phase. Such effects may be
important if the chemical reaction time is considerably smaller than
the time on which the local temperature and density changes, which
require that convection is much faster than chemistry. Chemical
equilibrium, in contrast, holds if the chemical reactions are much
faster than the hydrodynamic changes. Since photochemical processes
are irrelevant in inner substellar atmospheric regions, the above
ideas are concerned with the local thermo- and hydrodynamic condition,
or in other words, it can be considered in Lagrangian coordinates.
Hence, observation of CO and CH$_4$ for instance in Gl~229B do not
necessarily mean that the convective time scale is smaller than the
chemical reaction time for these molecules. It may very well be that a
convective cell of hot gas with its respective chemical equilibrium
composition travels upward and is detected in a considerable cooler
surrounding. However, the local conditions inside this cell are
appropriate for the presence of CO and the other gas compounds in
chemical equilibrium since they are not influence by the surrounding
nor by photo-processes yet. Hence, there are at least two reasons not
to expect chemical non-equilibrium: 1) the atmospheres are almost
entirely convective thereby providing a efficient mixing mechanism,
and 2) substellar atmospheres are very dense, hence, the chemical time
scales are very small due to high collision rates.

\paragraph{\underline{Dust formation}} 

First steps for understanding dust formation and for providing a
consistent description for substellar atmospheres are made (see
Sect.~\ref{sec:lsr}). So far, the extended model equation for dust
formation which include the drift effect have not be solved
consistently inside a model atmosphere. 

Note that the formation process of dust is a non-equilibrium process
since it takes place far away from thermodynamic equilibrium. The gas
needs to be substantially over-saturated, only then the phase
transition is kinetically possible.

The next crucial step may be the consideration of coagulation
processes. Coagulation is likely to play a role in the inner, drift
dominated atmospheric layers. Here the drift may cause grain-grain
collision and thereby an non-continuous, jump-like dust growth process
leading to much larger dust grains than we presently do predict.

\begin{figure*}
\vspace*{-2cm}
\centerline{\epsfig{file=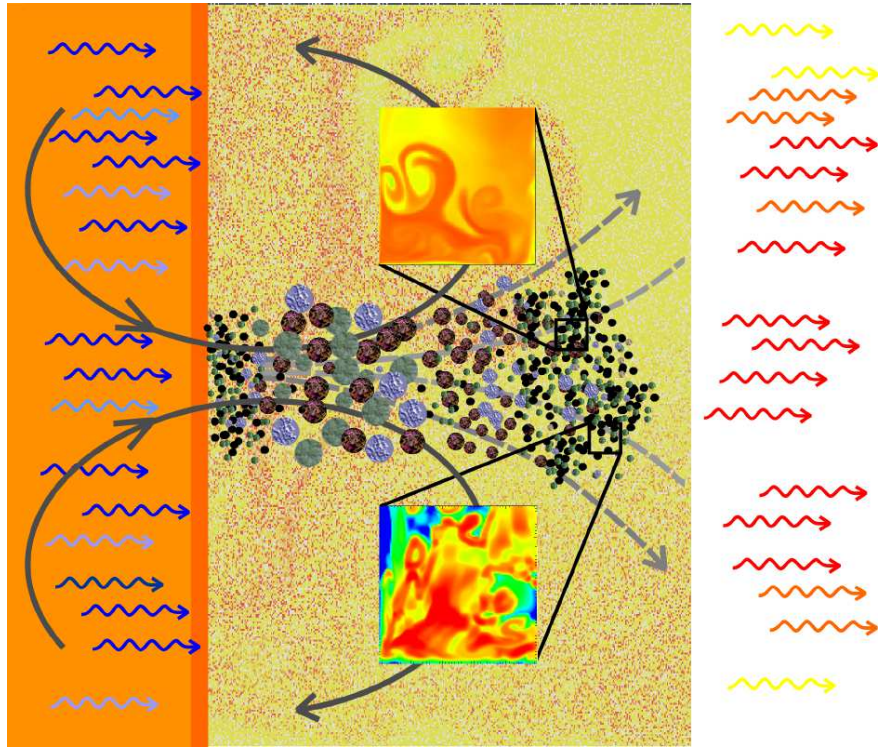, scale=0.35}}
\caption{Sketch of processes determining the appearance of a
substellar object. A strong and deep convection (closing arc) causes
convective overshoot (outward dashed arrows) which transports element
enriched material outwards. Inside such up-streams dust formation
takes place, appears as spots on the object's surface, and the formed
grains settle gravitationally. Radiation can only escape through holes
in the atmosphere resulting in a certain small scale variability while
the appearance of large dust clouds may cause the observable
photometric variability.}
\label{fig5}
\end{figure*}

\section{Summery}\label{sec:concl}

We have presented work on two entirely different scale regimes of
substellar atmospheres, one dealing with the observational large-scale
regime and the other dealing with the small-scale regime where
structure formations can be initiated. This allows us to cautiously
draw a picture which sketches processes and interactions most relevant
to rain and cloud formation in substellar
atmospheres. Figure~\ref{fig5} summarises the expected large scale
structure as well as the small scale structure formation due to
turbulence seen as a zoom in into the large scale structures of the
atmosphere.

First, the turbulent character of a substellar atmosphere does allow
the formation of grains in else-wise dust-hostile regions,
i.e. further inside the atmosphere. The constantly occurring wave
input from e.g. colliding convection cells causes the dust to form and
later to gather into larger and larger structure (clouds) which can
block radiation and cause backwarming due to the high opacity of dirty
dust grains.

Second, the dust formation does very likely occur inside elemental
rich, upstreaming gaseous material. Once first seed particles have
formed, they immediately start to sink into deeper, much denser
atmospheric layer. During this travel the particle grow to $100\mu$m
sized grains and speed up their fall as they get larger. The region of
seed formation is characterised by very small grains and a high metal
deficiency. The dust particles each their maximum size in deep layers,
just at the borderline of thermal stability.  What we know from Earth
rain we find also for brown dwarfs, the rain drops simply path through
some atmospheric regions before they hit the ground or evaporate as it
is in the case of brown dwarfs and giant gas planets.  Hence, there is
a distinct upper and lower boundary where the dust clouds are to be
expected. The upper borderline is related to th condition that enough
condensable elements must be mixed upwards, the lower borderline is
given by the condition of thermal stability. It is clear from
thermodynamic arguments, that the cloud layer sinks deeper into the
atmosphere as cooler the object is and visa versa. It is, hence, not
surprising that cooler substellar objects do appear more metal
deficient.

\begin{acknowledgements}
  This work has been supported by the \emph{DFG} (grants SE
  420/19-1\&2, Kl 611/7-1, Kl 611/9-1) and the {\it Berliner Programm
  zur F\"orderung der Chancengleichheit f\"ur Frauen in Forschung und
  Lehre}. Most of the literature search has been performed with the
  ADS system.

\end{acknowledgements}

\bigskip

\noindent
{\bf Burgasser:}
   Are the structures you show at later times an analogous "picture"
   of what you see on the surface of a dust-forming star / brown
   dwarfs? Have you considered rotation effects?  --  great talk

\smallskip
\noindent
 1) In case of true self-similarity, this could be true. However,
   I would hesitate to simply scale up the present results since they
   are obtained for a very different scale regime where major large
   scale processes like for instance convection have no direct
   influence on the local fluid dynamics, though they indirectly are
   present as driving mechanism for turbulence.

\noindent
2) Rotation effects are to be expected in the large-scale regimes
but not in the small-scale regimes investigated so far.

\medskip
\noindent
{\bf Jevremovic:} What is the size of the numerical grid?  What is the
time scale of transition from the adiabatic to the isothermal phase?

\smallskip
\noindent
1) The 2D simulations have a resolution of $128 \times 128$ grid
   cells, the 1D simulation 500 grid cells both covering a physical
   size of 500m.

\noindent
2) The local transition time scale is very small and it merely
depends on the dust formation time scale. The global time scale will
be much longer in a turbulent medium since areas can remain dust-free
longer than others. To give a rough estimate, duration from $2-20$\,s
are possible, depending also on the Mach number of the fluid.

\medskip
\noindent
{\bf Linsky:} Have you included the addition and subtraction of energy from the
   gas a result of condensation and evaporation processes? Will this feedback on the thermal structure of the atmosphere?

\smallskip
\noindent
1) Non of the dust simulations presented here need to include such
reaction energy effects as is has been discussed in (Woitke \& Helling
2003 A\&A 399, 297). We have shown that this effect can only amount to
a maximum heating of $\approx 3$\,K of any grain which we consider as
negligible under typical brown dwarf atmosphere conditions.

\noindent
2)The release of condensation heat during grain grow can be important
for the gas, possibly driving the convection as in the Earth
atmosphere. The abundance of high-temperature condensates in brown
dwarf atmospheres is smaller than that of water in the Earth
atmosphere. Therefore, we expect smaller effects.

\medskip
\noindent
{\bf Ludwig:} Which is the dominant spatial scale (relative to the
 scale on which turbulence is driven) of the dust resulting
 structures?  In how far do smaller scales influence the dominant dust
 structures?  Which is the minimum level of fluctuations necessary to
 significantly influence dust formation?

\smallskip
\noindent
1) This scale evolves over time from small ($\approx$ grid resolution
   $\approx 10^2$ cm in 2D) to large (size of test volume = 500
   m) according to our simulations.

\noindent
2) They profoundly influence the large-scale dust structures since the
   dust formation is initiated in the small-scales regimes.

\noindent
3) The smallest global Mach number studied so far is $M=0.2$ which
   determines the energy input at the largest scale. The local
   turbulent fluctuation does, however, fall into a regime of smaller
   Mach numbers. The quantitative character of structure formation is
   very similar to the $M=1$ regime, but the time scales are longer.

\medskip
\noindent
{\bf Steinacker:}
The present gas/dust structures are spatially very complicated, how
   do you handle the radiative transfer and especially the dust
   cooling?

\smallskip
\noindent
Proper radiative transfer is not included so far.  Present dust
 cooling is treated by a relaxation ansatz towards a constant
 temperature.  Multi-dimensional radiative transfer simulations do,
 however, suggest the support of such inhomogeneous media than a
 smoothing out of these structures (see e.g. Woitke el al. 2000 A\&A
 358, 665).

\medskip
\noindent
{\bf Dobler:} In your model the stratification quickly changes from
   adiabatic to isothermal, which is very different from what we have
   in the Earth's troposphere. What is the physics behind this
   difference?

\smallskip
\noindent
First, the optical properties of the considered particles are large in
brown dwarf atmospheres. We have to expect dirty,
i.e. non-transparent, dust particles in brown dwarfs which is very
different to what we have in the Earth's atmospheric layers. Second,
the ratio radiative/thermal energy of the gas is much larger in brown
dwarf atmospheres since it scales with $\sim T^3$ and brown dwarfs are
considerably warmer than the Earth.

\end{document}